\documentclass[pdflatex,sn-mathphys-num]{sn-jnl}% Math and Physical Sciences Numbered Reference Style
\usepackage{graphicx}%
\usepackage{multirow}%
\usepackage{amsmath,amssymb,amsfonts}%
\usepackage{amsthm}%
\usepackage{mathrsfs}%
\usepackage[title]{appendix}%
\usepackage{xcolor}%
\usepackage{textcomp}%
\usepackage{manyfoot}%
\usepackage{booktabs}%
\usepackage{algorithm}%
\usepackage{algorithmicx}%
\usepackage{algpseudocode}%
\usepackage{listings}%
\usepackage{amsmath}
\usepackage{array}
\usepackage{ amssymb }
\usepackage{braket}
\usepackage{qcircuit}
\usepackage{soul}
\usepackage{braket} 
\usepackage{relsize}
\theoremstyle{thmstyleone}%
\theoremstyle{thmstyletwo}%

\theoremstyle{thmstylethree}%

\begin{document}

\title{A computation of the covariance between two linear statistics for the Jellium model} %% Article title
\author{Pete Rigas}


%% Abstract
\abstract{
  We extend previous results providing an exact formula for the variance of a linear statistic for the Jellium model, a one-dimensional model of Statistical mechanics obtained from the $k \longrightarrow 0^{+}$ limit of the Dyson log-gas. For such a computation of the covariance, in comparison to previous work for computations of the log-gas covariance, we obtain a formula between two \textit{linear statistics}, given arbitrary functions $f$ and $g$ over the real line, that is dependent upon an asymptotic approximation of the Jellium probability distribution function from large $N$ deviations.}




\maketitle

 \label{sec1}
%% Labels are used to cross-reference an item using \ref command.

\section{Introduction}

\subsection{Overview}

\noindent Log-gases have attracted great attention from both the Mathematics and Physics communities, with several studies pertaining to, exact extremal statistics, and index distribution, for the Coulomb gas in one-dimension [3,4], connections with random-matrix theory [7], large deviation principles [10], characterizations relating to the one-dimensional plasma [6,11], classification of Gibbs measures [1], a central Limit theorem result [13], in addition to several conjectures and open questions [9]. To extend the analysis provided for an exact formula of the variance of a linear statistic in the Jellium model provided in [5], which can further add to the collection of work in the literature for approximating the behavior of \textit{linear statistics} for a wide variety of models, including the truncation of linear statistics for top eigenvalues of random matrices [8], as well as another result for a \textit{covariance measure} [2], which under certain assumptions can be used to obtain a \textit{generalized covariance measure}, we extend methods of a recent work contained in [5] to obtain a covariance for two \textit{linear statistics}, in the large, strictly positive $N$ limit.

\subsection{Paper organization}

\noindent In the following, we present objects from the Jellium model for defining the \textit{energy function}, and connection with quantities of the \textit{Riesz gas}. Before introducing the \textit{covariance formula} which is dependent upon two \textit{linear statistics} as parameters, we provide an overview of the rate function $\Psi$, which is a function of the linear statistic. Following the overview of the methods contained in [5] for obtaining the behavior of the Variance statistic for large $N$, we provide a computation of the covariance for two linear statistics in the Jellium model, which differs from a \textit{covariance measure} previously obtained for the log-gas in [2], which is instead dependent upon an effective saddle-point density action.

\subsection{Motivation}

\noindent Our motivation for studying closed-form representations of the Jellium covariance function largely comes from other models of Statistical Mechanics, several of which have previously been examined by the author [12]. In particular, regardless of whether interactions between point particles are defined over the line, or over an entire lattice, restricting the nearest neighbor interactions of points particles along copies of the real line allows for generalizations of saddle-point actions. Such actions are suitable for further generalizations of covariance functions for other models. As a general matter of fact, being able to write out closed form representations of correlation functions for models in Statistical Physics can help characterize different rates of decay for such functions. Depending upon the phase diagram of a model of interest, there can exist thresholds above, and below, which the model exhibits significantly different behaviors. In the simplified setting of copies of the real line, the Jellium model, as related to the Riesz gas, can exhibit the following possible ranges of behavior, including: (1), a vanishing value of the covariance, or (2), a strictly positive value of the covariance, over configurations of particles distributed over the real line. From nearest neighbor interactions that are encoded in Hamiltonians for other models of Statistical Mechanics besides the Riesz gas, and Jellium model, covariance functions supported over $\textbf{Z}^2$, or other lattices, can behave similarly in the two previously described categories as observed in covariance functions for the Jellium model; however, more complicated global behaviors supported over a lattice can emerge from interactions between nearest-neighbor lines of particles above or below a copy of the real line.

\subsection{Jellium objects}

\noindent We introduce the following quantities.

\bigskip

\noindent \textbf{Definition} \textit{1} (\textit{Riesz gas energy}). Introduce, for some $x_i$ on $\textbf{R}$,

\begin{align*}
      \mathscr{E}^k \big(x_1,  \cdots ,  x_i , \cdots , x_N \big) \equiv \mathscr{E} \equiv  \frac{A}{2} \overset{n}{\underset{i=1}{\sum}}   x^2_i + \alpha \text{ }  \mathrm{sgn} ( k )   \underset{i \neq j}{\sum }  \big| x_i - x_j \big|^{-k}       \text{, } 
\end{align*}

\noindent corresponding to the \textit{energy} of the \textit{Riesz gas}, for strictly positive parameters $A$ and $\alpha$, with,

\begin{align*}
     A   \sim \mathrm{O} \big( 1 \big) \text{, } \\  \alpha \sim \mathrm{O} \big( 1 \big)  \text{, }  \end{align*}

\noindent and sign function,

\[
\mathrm{sgn}(k ) =  \text{ } 
\left\{\!\begin{array}{ll@{}>{{}}l}        1  \text{ , }  \text{ }  x > 0 \text{ , }   \\  
     0    \text{ , }  \text{ }   x = 0     \text{, } \\- 1  \text{ , }  \text{ }  x < 0          \text{, } 
\end{array}\right.
\]

\noindent The energy function above, with the positions of each particle at any position $i$, under the conformal transformation,

\begin{align*}
T \big( x \big) = \frac{ax+b}{cx+d} \text{, } 
\end{align*}

\noindent for $ad -cb \neq 0$, which sends the position of each particle at $x_i$ to $\lambda_i$, from,

\begin{align*}
   T \big( x_i \big) = \frac{ax_i+b}{cx_i+d}  \mapsto \lambda_i     \text{, } 
\end{align*}

\noindent yields a distribution on eigenvalues of random matrices, given by,

\begin{align*}
       \mathscr{E}^k \big( T \big( x_1 \big) , \cdots ,  T \big( x_i \big)  , \cdots , T \big( x_N  \big)  \big) \equiv  \mathscr{E}^k \big( \lambda_1 , \cdots , \lambda_i , \cdots , \lambda_N  \big) \equiv \big( \mathscr{E}^{\prime} \big)^k \equiv  \frac{A}{2} \overset{n}{\underset{i=1}{\sum}}   \lambda^2_i \\ +  \alpha \text{ }  \mathrm{sgn} ( k )   \underset{i \neq j}{\sum }  \big| \lambda_i - \lambda_j \big|^{-k}       \text{, } 
\end{align*}

\noindent under the same choice of couplings $A$ and $\alpha$.

\bigskip

\noindent \textbf{Definition} \textit{2} (\textit{Riesz gas probability measure on random-matrix eigenvalues, a Gibbs-Boltzmann distribution}). From the \textit{energy} of \textbf{Definition} \textit{1}, introduce,

\begin{align*}
\textbf{P} \big( \lambda_1 , \cdots , \lambda_i , \cdots , \lambda_N  \big)  \equiv \frac{1}{Z^k_N} \mathrm{exp} \bigg(  -    \frac{ ( \mathscr{E}^{\prime})^k}{k_b T }    \bigg)        \text{, } 
\end{align*}

\noindent given the partition function,

\begin{align*}
   Z^k_N \big( \lambda_1 , \cdots , \lambda_i , \cdots , \lambda_N \big)  \equiv     Z^k_N \equiv      \underset{i : i \leq N}{\sum}        \bigg(  \frac{A}{2} \overset{n}{\underset{i=1}{\sum}}   \lambda^2_i + \alpha \text{ }  \mathrm{sgn} ( k )   \underset{i \neq j}{\sum }  \big| \lambda_i - \lambda_j \big|^{-k}                  \bigg)    \text{, } 
\end{align*}

\noindent normalizing $\mathrm{exp} \big[  -    \mathscr{E} \big( \lambda_i \big)     \big]$ so that $\textbf{P} \big( \lambda_1 , \cdots , \lambda_i , \cdots , \lambda_N  \big) $ is a probability measure, for some $N > 0$, Boltzmann constant $k_b$, and temperature $T>0$, with,

\begin{align*}
T \sim \mathrm{O} \big( 1 \big) \text{  }\text{ .  }
\end{align*}

\noindent \textbf{Definition} \textit{3} (\textit{Rescaling the energy function of the Gibbs-Boltzmann weight to obtain the probability measure of the Jellium model for} $k \equiv -1$). Introduce, $ Z^{-1}_N \big( \lambda_1 , \cdots , \lambda_i , \cdots ,\lambda_N \big) \equiv  Z^{-1}_N$, with,

\begin{align*}
   Z^{-1}_N \equiv             \underset{i : i \leq N}{\sum}        \bigg(  \frac{A}{2} \overset{n}{\underset{i=1}{\sum}}   \lambda^2_i + \alpha \text{ }  \mathrm{sgn} ( -1 )   \underset{i \neq j}{\sum }  \big| \lambda_i - \lambda_j \big|^{-1}                  \bigg)  \equiv  \underset{i : i \leq N}{\sum}        \bigg(  \frac{A}{2} \overset{n}{\underset{i=1}{\sum}}   \lambda^2_i - \alpha \text{ }    \underset{i \neq j}{\sum }  \big| \lambda_i - \lambda_j \big|^{-1}                  \bigg)    \text{, } 
\end{align*}

\noindent corresponding to the partition function of the \textit{Riesz gas probability measure} of \textbf{Definition} \textit{2} for $k \equiv -1$, which under the rescaling yields the following summation of \textit{energy} functions, taking the form,

\begin{align*}
 Z^{-1}_N \big( L_N \lambda_1 , \cdots , L_N \lambda_i , \cdots , L_N \lambda_N \big) \equiv \text{ }    \underset{i : i \leq N}{\sum}        \bigg(  \frac{A}{2} \overset{n}{\underset{i=1}{\sum}}   \big( L_N \lambda \big)^2_i - \alpha \text{ }   L_N \text{ } \underset{i \neq j}{\sum }  \big| \lambda_i - \lambda_j \big|^{-1}                  \bigg)                \text{ , } 
\end{align*}

\noindent hence providing the \textit{probability measure} of the Jellium model, 

\begin{align*}
\textbf{P}  \big( L_N \lambda_1 , \cdots , L_N \lambda_i , \cdots , L_N \lambda_N   \big)  \equiv  \widetilde{\textbf{P} } \big( \lambda_1 , \cdots , \lambda_i , \cdots , \lambda_N  \big)  \equiv \frac{1}{{Z^{-1}_N} } \mathrm{exp} \bigg(  -    \frac{(\mathscr{E}^{\prime} )^{-1}  }{k_b T }    \bigg)         \text{, } 
\end{align*}

\noindent for the rescaled Jellium \textit{energy function},

\begin{align*}
\mathscr{E}^{-1} \big( L_N \lambda_1 , \cdots ,  L_N \lambda_i , \cdots , L_N \lambda_N \big)     \equiv   \big(\mathscr{E}^{\prime}\big)^{-1} \equiv  \frac{A}{2} \overset{n}{\underset{i=1}{\sum}}   \big( L_N \lambda \big)^2_i - \alpha \text{ }   L_N \text{ } \underset{i \neq j}{\sum }  \big| \lambda_i - \lambda_j \big|^{-1}     \text{, } 
\end{align*}

\noindent given a suitable constant $L_N >0$ dependent upon the choice of $N$.

\bigskip

\noindent \textbf{Definition} \textit{4} (\textit{linear statistics}). A \textit{linear statistic} takes the form,

\begin{align*}
    s \equiv \frac{1}{N} \overset{N}{\underset{i=1}{\sum}} f \big( x_i \big)    \text{, } 
\end{align*}

\noindent for an arbitrary function $f(x)$ of the position $x$ on the real line, under the same choice of parameter $N$ provided in \textbf{Definition} \textit{2}.

\bigskip

\noindent \textbf{Definition} \textit{5} (\textit{covariance formula from the Gibbs-Boltzmann weight, } {[2]}). Introduce, for $\Lambda^N \subsetneq \textbf{R}$,

\begin{align*}
   \mathrm{Cov} \big( A   ,      B   \big) \equiv \underset{\Lambda^N}{\int}  \bigg(  \underset{1 \leq i \leq n}{\prod}  \mathrm{d} \lambda_i  \bigg)    \frac{1}{Z^{-1}_N} \mathrm{exp}\bigg(   -    \frac{(\mathscr{E}^{\prime} )^{-1}  }{k_b T } \bigg)  A \big( T \big( x_1 , \cdots , x_N \big) \big) B \big( T \big(   x_1 , \cdots , x_N \big) \big) \\ - \langle A \rangle \langle B \rangle    \text{, }
\end{align*}

\noindent corresponding to the \textit{covariance formula} between the \textit{linear statistics},  

\begin{align*}
  A      =   \frac{1}{N}  \overset{N}{\underset{i=1}{\sum}} f \big( x_i \big)  \text{, } 
\end{align*}

\noindent and,

\begin{align*}
   B     =    \frac{1}{N} 
    \overset{N}{\underset{i=1}{\sum}} g \big( x_i \big)    \text{, } 
\end{align*}

\noindent for two arbitrary functions $f(x)$ and $g(x)$, and $\Lambda \subsetneq \textbf{R}^N$, for the expectation,

\begin{align*}
    \langle \cdot \rangle_{A,\textbf{R}} \equiv   \langle \cdot \rangle_{A}   \text{, }
\end{align*}

\noindent against a test \textit{linear statistic} $A$.

\subsection{Overview of the Coulomb gas method}

\noindent In this section, we recount how the Gibbs-Boltzmann weight, adapted from the log-gas to the Jellium model with \textbf{Definition} \textit{3}, below. The choice of such a weight function is significant for the formula for the covariance that is obtained with an adaptation of the saddle-point method. As it will be shown, with the saddle-point method and a large deviations argument, the effective potential function $V^{\mathrm{eff}} \big( \cdot \big)$ imposes constraints on the large deviations approximation of the Jellium probability measure. Hence, one can obtain the desired representation for the covariance formula of the Jellium model. As a function of two arbitrary linear statistics, the covariance function is approximated with a suitable exponential whose power is proportional to the saddle-point action.

\bigskip

\noindent First, introduce a probability measure on the \textit{linear statistic}, with,

\begin{align*}
\mathcal{P} \big[  s  , N \big] \equiv  \underset{\Lambda^N}{\int}  \bigg( \underset{1 \leq i \leq n}{\prod} \mathrm{d}   x_i  \bigg)  \widetilde{\textbf{P}}  
 \big(   x_1 , \cdots ,  x_i, \cdots  x_N \big)  \delta_{s - \frac{1}{N} {\underset{i=1}{\sum}} f (  x_i  ) } 
\end{align*}

\noindent which under conformal transformations in \textbf{Definition} \textit{1}, is,

\begin{align*}
  \mathcal{P} \big[ T s  , N \big] \equiv  \underset{\Lambda^N}{\int} \bigg(  \underset{1 \leq i \leq n}{\prod} \mathrm{d} \big(  T \big( x_i \big) \big) \bigg)  \widetilde{\textbf{P}}  
 \big(  T \big( x_1 \big) , \cdots ,  T \big( x_i \big), \cdots  T \big( x_N \big) \big) \delta_{s - \frac{1}{N} {\underset{i=1}{\sum}} f (  T ( x_i  ) ) }   
 \end{align*}

 \begin{align*} \equiv  \underset{\Lambda^N}{\int}  \bigg( \underset{1 \leq i \leq n}{\prod} \mathrm{d} \lambda_i \bigg)  \widetilde{\textbf{P}}  
 \big( \lambda_1 , \cdots , \lambda_i , \cdots \lambda_N \big) \delta_{s - \frac{1}{N} {\underset{i=1}{\sum}} f ( \lambda_i ) }    \text{, } 
\end{align*}

\noindent which, upon rescaling the two interactions in the \textit{energy} function of the \textit{Jellium model}, has a mass of order, by a large deviation,

\begin{align*}
  \mathcal{P} \big[ s , N \big]  \approx \mathrm{exp} \big(      - N^3 \Psi \big( s \big)    \big)        \text{, } 
\end{align*}

\noindent where in the power of the exponent above, besides the $N^3$ scaling, is given by the rate function,

\begin{align*}
  \Psi \big( s \big)   \approx \frac{1}{2b} \big( s - \bar{s} \big)^2       \text{, } 
\end{align*}

\noindent about the minimum $s = \bar{s}$ of the rate function, with variance,

\begin{align*}
\mathrm{Var} \big[ s \big] \approx \frac{b}{N^3} \text{, } 
\end{align*}

\noindent for some positive parameter $b$. We proceed by summarizing components for each quantity in $\mathcal{P} \big[ s , N\big]$, in which approximations are introduced for,

\begin{align*}
     \delta_{s - \frac{1}{N} {\underset{i=1}{\sum}} f ( x_i ) }        \approx             N^3 \underset{\Gamma}{\int}   \frac{1}{2 \pi} \mathrm{exp} \big(  - \nu N^3       \big) \mathrm{d} \mu     \text{, } 
\end{align*}

\noindent and hence also for,

\begin{align*}
    \underset{\Lambda^N}{\int} \bigg(  \underset{1 \leq i \leq n}{\prod}  \mathrm{d} \lambda_i  \bigg)    \widetilde{\textbf{P}}  
 \big( \lambda_1 , \cdots , \lambda_i , \cdots \lambda_N \big)      \approx \frac{1}{Z^{-1}_N } \underset{\Gamma}{\int} \mathrm{d} \mu   \underset{\Lambda^N}{\int}   \bigg(  \underset{1 \leq i \leq n}{\prod}  \mathrm{d} \lambda_i  \bigg)        \text{ } \mathrm{exp}\bigg(   -    \frac{(\mathscr{E}^{\prime} )^{-1}  }{k_b T } \bigg)              \text{, }  \tag{\textit{*}}
\end{align*}

\noindent given $\Gamma \subsetneq \textbf{C}$ intersecting the imaginary axis of the $\mu$ plane. Given an arbitrary $f(x)$, the effective potential,

\begin{align*}
V^{\mathrm{eff}} \big( x \big) \equiv \frac{1}{2} x^2 + \mu f(x) \text{ } \text{ . } 
\end{align*}

\noindent For $\mu \equiv 0$, adopt the convention,

\begin{align*}
   Z^{-1}_N \big( \mu =  0 \big) = Z^{-1}_N      \text{ } \text{ . } 
\end{align*}

\noindent As a function of nonzero $\mu$, for large $N$ the hydrodynamic approximation the partition function appearing in $(\textit{*})$ has been shown to be of the form,

\begin{align*}
Z^{-1}_N \approx               \underset{\Lambda^N}{\int} \bigg( \underset{1 \leq i \leq n}{\prod} \mathrm{d} \lambda_i \bigg) \mathrm{d} \big( \rho \big( x \big) \big) \text{ }   \mathrm{exp} \bigg[   - N^3 \textbf{E}_{\mu} \big[ \rho(x)   ]  - N \underset{\mathrm{supp}}{\int}   \mathrm{d} x \rho(x) \mathrm{ln} \rho(x)  \bigg]          \delta_{\int \mathrm{d} x \rho(x)  -1}  \text{, }  \tag{\textit{**}}
\end{align*}

\noindent where the expectation of the density profile $\rho \big( x \big)$ appearing in the power of the exponential is,

\begin{align*}
      \textbf{E}_{\mu} \big[ \rho \big(x \big)  \big]   = \underset{\mathrm{supp}}{\int}  \bigg( \frac{x^2}{2} + \mu f \big( x \big) \bigg) \rho \big( x \big) \mathrm{d} x - \alpha \underset{\mathrm{supp}}{\int} \text{ }   \underset{\mathrm{supp}}{\int}  \rho \big( x \big) \rho \big( y \big)   \big| x - y \big| \mathrm{d} x \mathrm{d} y - \mu s   \text{, } 
\end{align*}

\noindent for the support of the profile over the real line,

\begin{align*}
\mathrm{supp} \cap \textbf{R} \neq \emptyset \text{, } 
\end{align*}

\noindent which admits the decomposition,

\begin{align*}
\mathrm{supp} \equiv \underset{1 \leq i \leq N}{\bigcap} \mathrm{supp}_i \text{, } 
\end{align*}

\noindent hence implying, under conformal transformations $T$, as those mentioned in \textbf{Definition} \textit{1}, 

\begin{align*}
   \underset{\mathrm{supp}}{\int} \mathrm{d} \big( \rho \big( T \big( x \big) \big) \equiv   \underset{1 \leq i \leq N}{\prod}  
 \underset{\mathrm{supp}_i}{\int} \mathrm{d} \big( \rho \big( T \big( x_i \big) \big) \text{ }  \text{ , }     \tag{$\mathrm{Supp}$}
\end{align*}

\noindent which is taken with respect to,

\begin{align*}
\textbf{E}_{\mu} \big[ \cdot \big] \text{ } \text{ . } 
\end{align*}

\noindent Altogether, under the assumption that, the effective action given on the LHS below,

\begin{align*}
     -    \frac{(\mathscr{E}^{\prime} )^{-1}  }{k_b T }  \approx N^3 S \big[ \rho \big( x \big) , \mu , \mu_0 \big]    \text{, } 
\end{align*}

\noindent for the action given by,

\begin{align*}
     S [ \rho \big( x \big) , \mu , \mu_0 ]  =    \underset{\mathrm{supp}}{\int}  \frac{x^2}{2} \rho \big( x \big) \mathrm{d} x - \alpha 
 \underset{\mathrm{supp}}{\int}  \text{ } \underset{\mathrm{supp}}{\int} \text{ } 
 \rho\big( x \big)  \rho \big( y \big) \big| x - y | \mathrm{d} x \mathrm{d} y  +   \mu \bigg(  \underset{\mathrm{supp}}{\int}  
 f \big( x \big) \rho \big( x \big) \mathrm{d} x - s \bigg) \\ +  \mu_0 \bigg(  \underset{\mathrm{supp}}{\int}   \rho \big( x \big) \mathrm{d} x - 1 \bigg)   \text{ , } 
\end{align*}

\noindent the expression for $\mathcal{P} \big[ s , N \big]$ in $(\textit{*})$ can be approximated with,

\begin{align*}
   \mathcal{P} \big[ s , N \big] \approx  \frac{1}{Z^{-1}_N }  \underset{\mathrm{supp}} 
 {\int} \mathrm{d} \big( \rho \big( x \big) \big) \underset{\Gamma}{\int}   \mathrm{d} \mu      \underset{\Gamma}{\int}   \mathrm{d} \mu_0 
 \text{ } \mathrm{exp} \bigg[  - N^3                 S \big[ \rho \big( x \big) , \mu , \mu_0 \big]         \bigg] \approx \bigg(  \underset{1 \leq i \leq N}{\prod} \underset{\mathrm{supp}_i} 
 {\int} \mathrm{d} \big( \rho \big( x_i \big) \big) \bigg) \\ \times \underset{\Gamma}{\int}   \mathrm{d} \mu      \underset{\Gamma}{\int}   \mathrm{d} \mu_0 
 \mathrm{exp} \bigg[  - N^3                 S \big[ \rho \big( x_i \big) , \mu , \mu_0 \big]         \bigg]   \text{ } \text{ . } \tag{\textit{***}}
\end{align*}

\noindent With the saddle-point method, the order of the unconstrained Jellium partition function is approximately,

\begin{align*}
  \frac{1}{Z^{-1}_N}  \approx          \frac{1}{\mathrm{exp} \big( \frac{2}{3} \alpha^2 N^3 \big)} \text{, } 
\end{align*}

\noindent for the rate function satisfying,

\begin{align*}
     \Psi \big( s \big)   \approx   \frac{2}{3} \alpha^2     \text{ } \text{ . } 
\end{align*}

\noindent Up to higher orders, the rate function above can also take the form,

\begin{align*}
\Psi \big( s \big) =         S \big[ \rho^{*}_{\mu} \big( x \big)  , \mu , \mu_0  \big] + \frac{2}{3} \alpha^2  \text{, } 
\end{align*}

\noindent for the effective saddle-point action, 

\begin{align*}
  S \big[ \rho^{*}_{\mu} \big( x \big)  , \mu , \mu_0  \big]  =   \underset{\mathrm{supp}}{\int}  \frac{x^2}{2} \rho^{*}_{\mu} \big( x \big) \mathrm{d} x - \alpha 
 \underset{\mathrm{supp}}{\int} \text{ } 
 \underset{\mathrm{supp}}{\int}  \rho^{*}_{\mu} \big( x \big)  \rho^{*}_{\mu} \big( y \big) \big| x - y | \mathrm{d} x \mathrm{d} y  + \mu \\ \times  \bigg(  \underset{\mathrm{supp}}{\int}  
 f \big( x \big) \rho^{*}_{\mu} \big( x \big)\mathrm{d} x - s \bigg)   \mu_0 \bigg(  \underset{\mathrm{supp}}{\int}   \rho^{*}_{\mu} \big( x \big)\mathrm{d} x - 1 \bigg)           \text{, }  \tag{\textit{****}}
\end{align*}

\noindent upon substituting (\textit{****}) into the power of the exponential given in $(\textit{***})$, namely replacing the saddle point action $S \big[ \rho^{*}_{\mu} \big( x \big)  , \mu , \mu_0  \big]$ with $S \big[ \rho_{\mu} \big( x \big)  , \mu , \mu_0  \big]$ yields the distribution,

\begin{align*}
     \mathcal{P} \big[ s , N \big] \approx  \frac{1}{Z^{-1}_N } \bigg(  \underset{1 \leq i \leq N}{\prod} \underset{\mathrm{supp}_i} 
 {\int} \mathrm{d} \big( \rho \big( x_i \big) \big)\bigg)  \underset{\Gamma}{\int}   \mathrm{d} \mu      \underset{\Gamma}{\int}   \mathrm{d} \mu_0 
 \text{ } \mathrm{exp} \bigg[  - N^3                 S \big[ \rho^{*}_{\mu} \big( x_i \big) , \mu , \mu_0 \big]         \bigg]   \tag{\textit{*****}}    \text{, } 
\end{align*}

\noindent which corresponds to the probability density function, given the support of each density $d \big( \rho \big( x_i \big) \big)$, each of which satisfies,

\begin{align*}
      \mathrm{supp}_i \cap \textbf{R} \neq \emptyset  \text{, } 
\end{align*}

\noindent for each $1 \leq i \leq N$. From such a closed expression for the rate function, the variance of a \textit{linear statistic} for large $N$ is extracted upon making the observation that,

\begin{align*}
    \bar{s} = \frac{1}{4 \alpha} \underset{[-2 \alpha , + 2 \alpha]}{\int} f \big( x \big) \text{ } \mathrm{d} x    \text{, } 
\end{align*}

\noindent is a minimum of $\Psi \big( s \big)$. Performing an expansion about the rate function about $s \equiv \bar{s}$ can be obtained by computing $\mu$ up to order $\mathrm{O} \big( \epsilon \big)$, for $\epsilon$ sufficiently small. The specific form of the rate function and effective saddle-point action are incorporated into the formula for the Jellium covariance for two \textit{linear statistics} obtained in the second section.

\section{Formalizing the computation for the Jellium covariance}

\noindent Below, we state the main result for the computation of the covariance. As emphasized previously, the computation of the covariance function for the Jellium model is dependent upon two linear statistics, which are introduced below. Straightforwardly, given the fact that each linear statistic is defined in terms of a summation over strictly positive $N$ measurements of sufficiently smooth functions $f$, and $g$, the covariance formula depends upon an integral over the support of the saddle-point action. Moreover, given the formula for interactions of the Riesz gas, the partition function appears as a normalization of the integral over the support of the saddle-point action. 

\bigskip

\noindent \textbf{Theorem} (\textit{computation of the Jellium covariance between two linear statistics}). For two \textit{linear statistics},

\begin{align*}
 A \equiv  A  \big( f,  x_1 , \cdots , x_i , \cdots , x_N \big)     =   \frac{1}{N}  \overset{N}{\underset{i=1}{\sum}} f \big( x_i \big)  \text{, } 
\end{align*}

\noindent and,

\begin{align*}
  B \equiv  B \big( g,  x_1 , \cdots , x_i , \cdots , x_N \big)      =    \frac{1}{N} 
    \overset{N}{\underset{i=1}{\sum}} g \big( x_i \big)    \text{, } 
\end{align*}

\noindent the Jellium covariance between $A$ and $B$ takes the form,

\begin{align*}
    \mathrm{Cov}^{\mathrm{Jellium}} \big( A   ,      B   \big)  \equiv    \frac{1}{N^2  }  \bigg( \underset{\mathrm{supp}} 
 {\int} \mathrm{d} \big( \rho \big( x  \big) \big) \underset{\Gamma}{\int}   \mathrm{d} \mu      \underset{\Gamma}{\int}   \mathrm{d} \mu_0 
 \text{ } \frac{1}{Z^{-1}_N} \mathrm{exp} \bigg[  - N^3                 S \big[ \rho^{*}_{\mu} \big( x \big) , \mu , \mu_0 \big]         \bigg] \\ \times         \bigg( \overset{N}{\underset{i=1}{\sum}} f \big(   x_i \big)  g \big(   x_i  \big) \bigg)   -  \bigg\langle  \overset{N}{\underset{i=1}{\sum}} f \big( x_i \big)   \bigg\rangle_f \bigg\langle \overset{N}{\underset{i=1}{\sum}} g \big(  x_i \big)  \bigg\rangle_g \bigg)  
   \text{ }             \text{, } 
\end{align*}

\noindent for some $N>0$, given $f$ and $g$ arbitrary.

\bigskip

\noindent In the following computation of the covariance, the conformal transformation $T$ which transforms to the position of Riesz gas particles to eigenvalues of some random matrix is related to the rate function, and hence, to the saddle-point action. The expectation with respect to the Jellium probability measure is related to an $n$-fold integral over independent copies of the sample space $\Lambda$, with $\Lambda^N \equiv \Lambda \times \cdots \times \Lambda$. By making use of such a representation for the expectation  $\langle \cdot, \cdots, \cdot \rangle_{\mathrm{Jellium}}$, the exponential of the saddle-point action is used an an approximation for the exponential of the Riesz gas Hamiltonian. Given two arbitrary linear statistics $f$ and $g$ introduced previously, the approximation which makes use of the exponential of the saddle-point action is incorporated into the desired representation for the covariance function.

\bigskip

\noindent \textit{Proof of Theorem}. Recall, from the previous section, that the approximation of the \textit{Riesz gas} energy for $k \equiv -1$, and hence of the Jellium model,

\begin{align*}
       -    \frac{(\mathscr{E}^{\prime} )^{-1}  }{k_b T }  \approx N^3 S \big[ \rho^{*}_{\mu} \big( x \big) , \mu , \mu_0 \big]      \text{, } 
\end{align*}

\noindent with the effective saddle-point density action $\rho_{\mu}^{*} \big( x \big)$, which appears in the contributions for the rate function,

\begin{align*}
   \Psi \big( s \big) =         S \big[ \rho^{*}_{\mu} \big( x \big)  , \mu , \mu_0  \big] + \frac{2}{3} \alpha^2   \text{, } 
\end{align*}

\noindent as well as the constant $\frac{2}{3} \alpha^2$. From previously implemented techniques for extracting the behavior of the variance of a single \textit{linear statistic} in the large $N$ limit, applying the conformal transformation,

\begin{align*}
T \big( x \big) = \frac{ax+b}{cx+d} \text{, } 
\end{align*}

\noindent provided in \textbf{Definition} \textit{1}, to each position $x_i$, of the \textit{energy} function of the \textit{Riesz gas} introduced in \textbf{Definition} \textit{1}, for $k \equiv -1$, given in \textbf{Definition} \textit{2}, yields,

\begin{align*}
\mathscr{E}^k \big( T \big( x_1 , \cdots , x_i , \cdots , x_N \big) \big) \equiv \mathscr{E}^k \big( T \big( x_1 \big) , \cdots ,  T \big( x_i \big)  , \cdots , T \big( x_N  \big)  \big) \\ \equiv  \mathscr{E}^k \big( \lambda_1 , \cdots , \lambda_i , \cdots , \lambda_N  \big) \text{, } 
\end{align*}

\noindent corresponding to the \textit{Riesz gas energy function}, under conformal $T$, for $k \equiv -1$. From the probability measure introduced in \textbf{Definition} \textit{3}, the quantity,

\begin{align*}
 \langle x_1 , \cdots , x_i , \cdots , x_N \rangle_{\mathrm{Jellium}}   \equiv    \underset{\Lambda^N}{\int} \bigg( \underset{1 \leq i \leq n}{\prod}  \mathrm{d} x_i \bigg)    \frac{1}{Z^{-1}_N}      \mathrm{exp} \bigg(     -    \frac{(\mathscr{E}^{\prime} )^{-1}  }{k_b T }      \bigg) \end{align*} 
 
 \noindent which can be approximated with,
 
 \begin{align*}
 \underset{\Lambda^N}{\int} \bigg( \underset{1 \leq i \leq n}{\prod} \mathrm{d} x_i \bigg)      \underset{\Gamma}{\int}   \mathrm{d} \mu      \underset{\Gamma}{\int}   \mathrm{d} \mu_0 
 \text{ } \frac{1}{Z^{-1}_N}     \mathrm{exp} \bigg[  - N^3                 S \big[ \rho^{*}_{\mu} \big( x \big) , \mu , \mu_0 \big]         \bigg]    \approx         \underset{1 \leq i \leq N}{\prod} \underset{\mathrm{supp}_i}{\int} \mathrm{d} \big( \rho \big( x_i \big) \big) \underset{\Gamma}{\int}   \mathrm{d} \mu       \underset{\Gamma}{\int}   \mathrm{d} \mu_0 \\ \times  
 \frac{1}{Z^{-1}_N}    \mathrm{exp} \bigg[  - N^3                 S \big[ \rho^{*}_{\mu} \big( x_i \big) , \mu , \mu_0 \big]         \bigg]   \text{, } 
\end{align*}

\noindent for the corresponds to the Jellium expectation taken across all $x_i$ up to $N$, for the \textit{energy function},

\begin{align*}
(\mathscr{E}^{\prime} )^{-1} \big( x_1 , \cdots , x_i , \cdots , x_N \big) \equiv  (\mathscr{E}^{\prime} )^{-1} \text{, } 
\end{align*}

\noindent which satisfies, given support over $\textbf{R}$ for the base probability measure $\widetilde{\textbf{P}} \big[ \cdot \big]$,

\begin{align*}
 \langle \cdot \rangle_{\mathrm{Jellium}}   \equiv   \langle \cdot \rangle_{\widetilde{\textbf{P}}, \textbf{R}}  \equiv   \langle \cdot \rangle     \text{ } \text{ . } 
\end{align*}

\noindent Proceeding, under conformal $T$ as introduced above applied to each $x_i$ as mentioned previously, from the ensemble average of each $T \big( x_i \big)$, the covariance between the two \textit{test statistics} can readily be formed, which coincides with,

\begin{align*}
  \langle x_1 , \cdots , x_i , \cdots , x_N \rangle_{\mathrm{Jellium}}   =     \bigg(\frac{1}{N^2}  \overset{N}{\underset{i=1}{\sum}} f \big(   x_i \big)  g \big(   x_i  \big) \bigg)   -  
 \frac{1}{N^2} \bigg\langle  \overset{N}{\underset{i=1}{\sum}} f \big( x_i \big)   \bigg\rangle_f \bigg\langle \overset{N}{\underset{i=1}{\sum}} g \big(  x_i \big)  \bigg\rangle_g      \text{, } 
\end{align*}

\noindent which, under the previous identification for  $\langle \cdot \rangle_{\mathrm{Jellium}}$, is equivalent to,

\begin{align*}
\frac{1}{N^2} \bigg(         \underset{1 \leq i \leq N}{\prod} \underset{\mathrm{supp}_i}{\int} \mathrm{d} \big( \rho \big( x_i \big) \big) \underset{\Gamma}{\int}   \mathrm{d} \mu      \underset{\Gamma}{\int}   \mathrm{d} \mu_0 
 \text{ } \frac{1}{Z^{-1}_N}     \mathrm{exp} \bigg[  - N^3                 S \big[ \rho^{*}_{\mu} \big( x_i \big) , \mu , \mu_0 \big]         \bigg]     \bigg(\frac{1}{N^2}  \overset{N}{\underset{i=1}{\sum}} f \big(   x_i \big)  g \big(   x_i  \big) \bigg) \\ -    \bigg\langle  \overset{N}{\underset{i=1}{\sum}} f \big( x_i \big)   \bigg\rangle_f \bigg\langle \overset{N}{\underset{i=1}{\sum}} g \big(  x_i \big)  \bigg\rangle_g           \bigg)         \text{ } \text{ . }
\end{align*}

\noindent Under suitable, conformal $T \big( \cdot \big)$ sending each position $x_i$ as input to a corresponding eigenvalue $\lambda_i$, the expression obtained above can be written as,

\begin{align*}
  \langle  T \big( x_1 , \cdots , x_i , \cdots , x_N \big) \rangle   \equiv    
 \langle T \big( x_1  \big) , \cdots , T \big( x_i \big)  , \cdots , T \big(  x_N \big) \rangle           \text{, } 
\end{align*}

\noindent which is in turn equivalent to,

\begin{align*}
  \frac{1}{N^2} \bigg(  \bigg(        \underset{1 \leq i \leq N}{\prod} \underset{\mathrm{supp}_i}{\int} \mathrm{d} \big( \rho \big( T \big( x_i \big) \big) \big) \bigg)  \underset{\Gamma}{\int}   \mathrm{d} \mu      \underset{\Gamma}{\int}   \mathrm{d} \mu_0 
 \text{ } \frac{1}{Z^{-1}_N}     \mathrm{exp} \bigg[  - N^3                 S \big[ \rho^{*}_{\mu} \big( T \big(  x_i \big)  \big) , \mu , \mu_0 \big]         \bigg]   \\ \times   \bigg(\frac{1}{N^2}  \overset{N}{\underset{i=1}{\sum}} f \big(  T \big(  x_i \big)  \big)  g \big( T \big( x_i  \big) \big) \bigg) -   \bigg\langle  \overset{N}{\underset{i=1}{\sum}} f \big( T \big(  x_i \big) \big)   \bigg\rangle_f \bigg\langle \overset{N}{\underset{i=1}{\sum}} g \big( T \big(  x_i \big)  \big)  \bigg\rangle_g           \bigg)             \text{, } 
\end{align*}

\noindent exhibiting that the desired formula for the Jellium covariance holds, under the observation, provided in ($\mathrm{Supp}$), that, 

\begin{align*}
      \underset{\mathrm{supp}}{\int} \mathrm{d} \big( \rho \big( T \big( x \big) \big) \equiv   \underset{1 \leq i \leq N}{\prod}  
 \underset{\mathrm{supp}_i}{\int} \mathrm{d} \big( \rho \big( T \big( x_i \big) \big)  \text{, } 
\end{align*}

\noindent and also that,

\begin{align*}
\underset{N \longrightarrow + \infty}{\mathrm{lim}} \langle \rho_N \big( x \big) \rangle = \rho \big( x \big)  \text{, } 
\end{align*}

\noindent which corresponds to the large $N$ limit of the density profile, which is given by,

\begin{align*}
\langle \rho_N \big( x \big)  \rangle \equiv \frac{1}{N} \bigg\langle 
 \overset{N}{\underset{i =1}{\sum}}    \delta \big( x - x_i \big) \bigg\rangle   \text{, } 
\end{align*}

\noindent corresponding to a summation over delta functions at each $x_i$. Hence we obtain a formula for the Jellium covariance that is a function of the two \textit{test statistics}, which are respectively given by arbitrary $f$ and $g$, from which we conclude the argument. \boxed{}

\section{Conclusion}

\noindent In this paper, we demonstrate how the computation of the covariance between two test statistics can be obtained for the Jellium model. As an adaptation of previous work that was extensively discussed in the Introduction and afterwards, the computation of the Jellium covariance function was dependent upon: (1) introducing a suitable energy functional for encoding interactions in the Riesz gas; (2) defining, from the Riesz gas energy functional, a probability measure normalized in the partition function of the model; (3) manipulating a probability measure on eigenvalues of random matrices, also obtained from the Riesz gas probability measure. Equipped with these quantities, we demonstrate, upon fixing two test statistics, that: (1) a large deviations principle, for approximating a probability functional with the exponential of the rate function, (2) saddle-point method, for expanding upon the method initially provided by Flack et al,  and (3) suitable conformal transformation, for transforming the positions of point particles of the Riesz gas to eigenvalues of a random matrix, together can be used to obtain the desired formula for the Jellium covariance function between two linear statistics. In comparison to other covariance functions which can be readily employed for hypothesis testing, and related statistical computations, the covariance function obtained for the Jellium function in this work not only takes into account physical characteristics of the Riesz gas, but also connections with fluctuations in the energy landscape through assumptions placed on the effective potential function, $V^{\mathrm{eff}} \big( \cdot \big)$.




\section{Declarations}

\subsection{Ethics approval and consent to participate}

The author consents to participate in the peer review process.

\subsection{Consent for publication}

The author consents to submit the following work for publication.

\subsection{Availability of data and materials}

Not applicable

\subsection{Conflict of interest}

The author declares no competing interests.

\subsection{Funding}

There are no funding sources.

\nocite{*}
\bibliography{sn-bibliography}% common bib file

\end{document}